\documentclass[fleqn,twoside]{article}
\usepackage{espcrc2}

\readRCS
$Id: espcrc2.tex,v 1.2 2004/02/24 11:22:11 spepping Exp $
\usepackage{graphicx}
\usepackage[figuresright]{rotating}

\newcommand{\AmS}{{\protect\the\textfont2
  A\kern-.1667em\lower.5ex\hbox{M}\kern-.125emS}}

\title{SCES `08 ~-~ concluding remarks }

\author{Suchitra E. Sebastian\address{Cavendish Laboratory, J J Thomson Ave, Cambridge CB3 0HE, UK} %
        \thanks{Suchitra E. Sebastian acknowledges financial support from Trinity College (Cambridge University), Royal Society conference grant, and the Institute for Complex Adaptive Matter.}, 
                C. Morais Smith\address{Institute for Theoretical Physics, University of Utrecht, Leuvenlaan 4, 3584 CE Utrecht, The Netherlands}%
        \thanks{C. Morais Smith acknowledges partial financial support from the Netherlands Organization for Scientific Research (NWO) and from the National Science Foundation under Grant No. NSF 
        PHY05-51164.}}
       
\begin{document}

\begin{abstract}
\vspace{1pc}
\end{abstract}

\maketitle

\section{INTRODUCTION}

This year's SCES conference has been as usual, memorable - not in small part because of the idyllic surroundings in the picturesque fishing town of Buzios, Brazil. We would especially like to thank the organisers, Elisa Saitovitch and Mucio Continentino for the sterling conference organisation. A breadth of exciting topics were discussed at this year's SCES, yet limitations of time mean that we are unable to do justice to them all. With a view to our particular areas of condensed matter specialisation and interest, we have chosen to focus on the area of novel materials - both those that occur in the solid state, and those that are artificially created.

The narrative of novel phase emergence in the vicinity of phase instabilities has now been central to strongly correlated electron systems for more than a decade. One of the first reasons for the attraction of condensed matter physicists to phenomena in the vicinity of a (nearly) continuous phase transition at zero temperature - known as a Quantum Critical Point (QCP) - was the possible appearance of universal behaviour where microscopic details of electrons and structures are bypassed in favour of macroscopic patterns \cite{Hertz1,Millis1}. This year's SCES has proved exciting in the array of unconventional phenomena discovered both in novel systems, and by the renewed investigation of age-old systems, arguably in the vicinity of QCPs. From heavy fermion systems, to cuprate superconductors, and in a new twist iron pnictide superconductors - some questions remain: just how similar or different are correlated phenomena in these systems? Further, how ubiquitous are ultra-strongly correlated effects such as the fractional quantum Hall effect (QHE), and can cold atom systems mimic such correlated phases? 

\section{HEAVY FERMION SUPERCONDUCTORS}

The puzzle of superconductivity in magnetic metals first came to the fore with the discovery of superconductivity in the heavy fermion systems UBe$_{13}$ \cite{Bucher1,Ott1} and CeCu$_2$Si$_2$ \cite{Steglich1} over a quarter of a century ago, a finding that was initially received amidst considerable incredulity. Soon, however, similar phenomena were discovered in a broad array of heavy fermion materials, revealing a pattern of superconductivity potentially mediated by magnetic interactions. The concept of novel phases mediated by enhanced interactions at a QCP came to the fore at this juncture and rapidly became ubiquitous.  It is intriguing, however, that despite the superficially similar fashion in which these materials were initially thought to behave, various aspects of the physics of these `model' systems continue to baffle.

\subsection{The 115 family}
The family of 115 heavy fermion systems perhaps constitutes the prototypical class of magnetic heavy fermion superconductors. Yet materials within this family continue to surprise.

\subsubsection{CeCoIn$_5$}

Ambient pressure superconductivity in the low-dimensional heavy fermion system CeCoIn$_5$ \cite{Petrovic1} followed the discovery of pressure-induced superconductivity in its three-dimensional (3D) analogue CeIn$_3$ \cite{Mathur1} a decade ago. The fact that unanswered questions continue to swirl thick and fast almost a decade after the discovery of superconductivity in CeCoIn$_5$ is a testament to the complexity of physics underlying an apparently straightforward case of magnetic interaction mediated superconductivity. A prominent debate at SCES this year pertained to whether or not a Fulde-Ferrell-Larkin-Ovchinnikov (FFLO) phase (i.e. a superconducting pairing state between Zeeman exchange split parts of the Fermi surface) \cite{Larkin1,Fulde1} is realised at high magnetic fields in CeCoIn$_5$. A new phase within the superconducting phase appears at high magnetic fields, and one interpretation is that this phase is the realisation of an FFLO \cite{Matsuda1}. 

Latest results of vortex lattice imaging studies as a function of magnetic field in CeCoIn$_5$ were presented, indicating a departure from Ginzburg Landau physics \cite{Bianch1}. Yet the case for the realisation of an FFLO phase at high magnetic fields is nebulous. Experimental results under pressure were presented to make the case for an FFLO ground state: the suppression of magnetism by the application of pressure was shown to result in an increased novel phase region, supporting the case for an FFLO phase \cite{Nicklas1}. However, recent results of high magnetic field neutron diffraction experiments challenge this notion of an FFLO. Evidence of long range antiferromagnetic order is found to be associated with the novel phase region of superconductivity at high magnetic fields; the invariance of the observed Cooper pair momentum with magnetic field appears to be inconsistent with a potential FFLO state \cite{Kenzelmann1}.

Of additional interest in the family of the 115's is their increasing similarity to the Cuprate family of high $T_c$ superconductors. In this case, commonalities appear to run more than skin deep. Not only does superconductivity appear to be related to magnetic interactions, but both evidence a form of density wave order, potentially antiferromagnetism only at high magnetic fields, perhaps suggesting a unique form of associated superconductivity. Further experiments to detect whether high-field antiferromagnetic phase is confined to the superconducting region may serve to shed light on this mystery.

\subsubsection{PuCoGa$_5$}
Another member of the 115 family in which superconductivity has recently been discovered is PuCoGa$_5$, with $T_c$ $\sim$ 18~K \cite{Sarrao1}. Spin susceptibility measurements that yield finite values at the lowest temperature have established a dirty d-wave form of the superconducting wavefunction in this material \cite{Curro1}. However, a spirited debate at this year's SCES pertained to whether unconventional superconductivity in this material is mediated by electron-phonon interactions or a form of magnetic interaction such as spin exchange between the magnetic lattice and the metallic environment.

While superconductivity in PuCoGa$_5$ develops out of a Pauli paramagnetic state, it is not entirely clear as to its proximity to a magnetic instability. The presence of spin-disorder scattering and local-moment behaviour that appears to be linked to the appearance of superconductivity have been cited as potential evidence for magnetically mediated superconductivity. As we heard in this year's SCES, however, theoretical work by Caicuffo et al. \cite{Caicuffo1} that models results of irradiation experiments on PuCoGa$_5$ served to caution us that nodal superconductivity does not necessarily imply magnetically mediated superconductivity, in fact most experimental features could be explained by means of an electron phonon mechanism within the Ehliashberg theory. A rebuttal of this point of view came from Piers Coleman, who instead proposed a different explanation in which spins constitute the fabric of exchange rather than the glue \cite{Flint1}. In this proposal, Kondo spin quenching and superconductivity develop simultaneously in a composite pairing mechanism involving Kondo spin exchange. Further experiments are required to distinguish between alternate mediating mechanisms of superconductivity. The variety of proposed mechanisms for unconventional superconductivity in PuCoGa$_5$ reveals yet again the complex physics arising from materials diversity within the 115 family of superconductors that have long been considered a model for magnetic interaction mediated superconductivity.

\subsection{New Heavy Fermion Superconductors}

\subsubsection {$\beta$-YbAlB$_4$}

A new entry into the class of $f$-electron superconductors was $\beta$-YbAlB$_4$. While following the theme of proximity to a phase instability, $\beta$-YbAlB$_4$ broke new ground in constituting the first known Yb-based $f$-electron superconductor, with $T_c\sim$ 80 mK \cite{Nakatsuji1}. It had remained another mystery of heavy fermion superconductors as to the prevalence of superconductivity in systems based on Ce (4f$^1$), but a singular absence in systems based on its single hole analogue Yb (4f$^{13}$). The discovery of $\beta$-YbAlB$_4$ goes some way in solving this mystery, and appears to follow the narrative of Ce-based heavy fermion superconductors: proximity to an antiferromagnetic instability. This remarkable discovery required measurements on ultra pure single crystals of $\beta$-YbAlB$_4$, and an experimental tour de force involving challenging low temperatures - in fact, far more stringent conditions required to observe superconductivity than in other $f$-electron families of materials. However, while proximity to a phase instability seems clear from the specific heat, magnetic susceptibility, and transport behaviour of $\beta$-YbAlB$_4$, the unconventional transport scaling behaviour renders unclear the nature of the neighbouring phase instability. Another mystery pertains to the far lower superconducting energy scale in this material as compared to Ce-based heavy fermion superconductors, possibly related to the size of antiferromagnetic interactions, the Fermi surface topology \cite{OFarrell}, or indeed, the nature of the neighbouring phase instability. The study of pressure induced magnetisation in $\beta$-YbAlB$_4$ may shed light on some of these puzzles, as will more experiments to probe the nature and symmetry of superconductivity in this material.

\subsubsection {NpPd$_5$Al$_2$}

An exciting new $f$-electron superconductor discussed at SCES this year was NpPd$_5$Al$_2$. A particularly intriguing aspect of this discovery was the accidental fashion in which this material was grown, resulting from an attempt to grow NpPd$_3$ single crystals out of Pb flux in Al$_2$O$_3$ crucibles. As it turned out, NpPd$_5$Al$_2$ was found to superconduct at 4.9~K \cite{Aoki1}. On the face of it, this material appears to carry some of the trademarks seen in unconventional $f$-electron superconductors. Preliminary experiments of specific heat, susceptibility, and the upper critical field performed on this material suggest potential $d$-wave singlet superconductivity in the paramagnetic limit. Currently, possibilities to explain the physics in this system are boundless. A potentially finite momentum superconducting state like CeCoIn$_5$, or composite pairing state have been suggested, yet further experiments alone will reveal as yet unexplored possibilities beyond the realm of familiar models.

Even as more members are added to the category of heavy fermion superconductors, the notion of a `universal QCP driving materials' properties appear more of a chimera than ever. Given the unexplained mysteries evidenced both in the familiar 115 family of compounds, and new families of heavy fermion superconductors, it is more likely than not that phenomena outside the confines of simple theories of quantum criticality are likely to emerge with more careful study measurements, and a broader scope of materials families under exploration. 

\section{FERROMAGNETIC SUPERCONDUCTORS}

The very notion of superconductivity existing in an itinerant ferromagnet was initially treated with disbelief, until its experimental discovery in UGe$_2$ \cite{Saxena1}. The overarching notion of superconductivity in close proximity to a near-continuous phase instability (in this case ferromagnetic) thought to underlie this phenomena motivated the discovery of more ferromagnetic superconductors such as URhGe \cite{Aoki2}. With new materials and more measurements, indeed, come unexpected findings that may not fit the simple narrative of QCP's, but may birth new discoveries in themselves.

\subsection{URhGe}

Interestingly enough, the case of UGe$_2$, where the narrative of novel phases in proximity to a QCP largely originated, has since been found to be more complex than first thought. The superconducting dome in UGe$_2$ lies in the vicinity of both a transition from a paramagnetic to ferromagnetic state, and a transition between two different ferromagnetic states - neither of these phase instabilities are thought to be continuous. So too the case of URhGe at ambient pressure and low magnetic fields. However, a new development arose with the discovery of unconventional superconductivity at high magnetic fields in the vicinity of a metamagnetic transition in URhGe \cite{Levy1}. Drawing on the theme of novel phases mediated in the vicinity of a continuous instability, the novelty of this discovery was in the different class of instability that was probed, further enabling tuning in a two-dimensional (2D) angular plane instead of along a one-dimensional (1D) axis. New measurements presented at SCES this year probed the enhancement in effective mass via transport measurements of the $A$-coefficient in URhGe \cite{Levy2,Miyake1}, thereby accessing the evolution of fluctuations in the vicinity of high field superconductivity. The striking finding from these experiments is a maximum in the $A$-coefficient that coincides with the peak of the superconducting dome, consistent with the notion of enhanced interactions in a quantum critical region that mediate unconventional superconductivity. The case of re-entrant superconductivity in URhGe appears to be a rarity in that it closely follows the simple theoretical description of superconductivity near a QCP, in this case terminating a plane of first order transitions - in fact branching off from a possible tricritical point \cite{Levy3}. URhGe is perhaps a model system where microscopic measurements may be used to probe the potential divergence of length-scales in the vicinity of a QCP. Further experiments of interest will no doubt be direct Fermi surface measurements to trace the enhancement in effective quasiparticle mass, and additionally, experiments that directly probe the superconducting wavefunction in this material, which could then be compared and contrasted with the case of UGe$_2$.

\subsection{New Ferromagnetic Superconductors}

\subsubsection{UCoGe}
As we have seen, there exists a breadth of different possibilities for the physics of phases mediated at a quasi-continuous instability even within the category of ferromagnetic superconductors. New members in this category, therefore, provide an excellent opportunity for further exploration. A new material we heard about at SCES this year was UCoGe \cite{Huy1}, a ferromagnetic superconductor in the manner of pressure-tuned UGe$_2$ and ambient pressure URhGe. Measurements of magnetisation, transport, thermal expansion, and specific heat reveal ferromagnetism at $T_c = 3~K$ that coexists with superconductivity at 0.8~K, suggesting that this material lies along the pressure-tuning axis, with its location to the left of the dome maximum - lying between UGe$_2$ to the left of the superconducting dome onset, and URhGe to the right of the superconducting dome maximum. Despite the similarities between these materials, however, they in fact display subtly different forms of magnetism - the magnetic transition is in the longitudinal moment in UGe$_2$, in the transverse moment in URhGe, and metamagnetism appears to be absent in UCoGe. Future experiments of mass enhancement and microscopic aspects of the magnetic phase transition in UCoGe may reveal deeper complexities that potentially underlie phase space in the vicinity of superconductivity in this material.

\subsubsection{CeFeAs$_{1-x}$P$_x$O}
Another material that has been discovered to lie close to a ferromagnetic QCP is an alloy of CeFeAsO and CeFePO \cite{Bruning1}. While CeFePO constitutes an extremely heavy fermion system ($\gamma \sim$ 1000 mJ/mol K$^2$) with no evidence of magnetic ordering, CeFeAsO exhibits antiferromagnetism associated with the lattice of local Ce moments at $T_n \sim$ 3.8 K, and $\gamma \sim$ 60 mJ/mol K$^2$. It has therefore been suggested that a ferromagnetic instability may lie partway between these two systems, and may be accessed by substituting As for P to yield CeFeAs$_{1-x}$P$_x$O.  The investigation of this material showed considerable foresight, considering that related members of this family of materials later formed parent systems of the recently discovered pnictide high $T_c$ superconductors. Much awaited experiments would involve further tuning of phase space to access this ferromagnetic phase instability in clean single crystals and investigating the possible emergence of unconventional superconductivity. 

\section{IRON PNICTIDE SUPERCONDUCTORS}
Arguably the high point of condensed matter discoveries this year was that of the new family of iron based high temperature superconductors. Interestingly enough, the discovery was made by a chemist Hideo Hosono, with the original goal of evaluating low-dimensional candidate materials for their potential as magnetic semiconductors. First signs of the imminent breakthrough came in 2006 and 2007 with the discovery of intrinsic superconductors LaFePO ($T_c \sim 4$K) \cite{Kamihara1} and LaNiPO ($T_c \sim 3$K) \cite{Tegel1}. In 2008, a significant advance was made with the finding of $T_c \sim 33$~K superconductivity in LaFeAsO doped with F \cite{Kamihara2}, to be followed up by the discovery of superconductivity on doping related materials such as REOFeAs (RE = La, Ce, Pr, Nd, Sm) and AFe$_2$As$_2$ (A = Ca, Sr, Ba, Eu) \cite{Norman1}. Parent materials in these families of compounds are typically antiferromagnets with relatively high temperature spin density wave transitions $T_{SDW} \sim$ 100 K, which upon doping or the application of pressure, achieve superconducting temperatures as high as $T_c \sim$ 55 K \cite{Ren1}. 

In a bid to bring to bear existing theoretical pictures on this new family of superconductors, the question has been posed as to whether the parent materials of these superconducting compounds are Mott insulators like the Cuprate family of superconductors, or in contrast, are closer to itinerant metals. Optical studies presented at SCES reveal a sharp drop in scattering rate and plasma frequency at the magnetic transition in these materials, providing robust evidence for an itinerant system gapped by a spin density wave resulting in the loss of the majority of carriers \cite{Hu1}. Results of quantum oscillation experiments presented at the meeting also reveal a very small remnant Fermi surface due to reconstruction by a spin density wave, indicating itinerant character \cite{Sebastian1}. A puzzle for theorists has been how to incorporate this itinerant character into a magnetic model of local exchange interactions constructed to understand the electronic structure in this system. One intriguing possibility presented at this conference was a `traffic light' model where J1-J2 exchange interactions are combined with itinerant electrons much in the fashion of traffic lights regulating the flow of traffic \cite{Uemura1}. The symmetry of the superconducting wavefunction in these materials is currently the subject of intense debate: detailed microscopic measurements on single crystals of improved purity will no doubt be crucial to progress on this front. 

\section{CUPRATE HIGH-$T_{C}$ SUPERCONDUCTORS}

A discussion of correlated electron systems would be incomplete without mentioning cuprate high-Tc superconductors. An understanding of unconventional superconductivity in these materials poses a conundrum that will hopefully be solved within the next few decades. There have been some important recent developments in the understanding of these materials, for instance, relating to the enigmatic pseudogap phase. We briefly touch on a few new experimental developments here.

Scanning tunnelling microscopy (STM) results from the group of
Yazdani suggest that high-$T_c$ superconductivity occurs in two steps:
at $T_{P}$ incoherent Cooper pairs are formed and at $T_c$ these preformed pairs break the U(1) gauge symmetry and reach phase coherence. It has been proposed that this intermediate temperature $T_{P}$ between $T^*$ and $T_c$ is closely related to the superconducting gap $\Delta$, given that $2 \Delta / k_B T_{P} \approx 8$ \cite{Yazdani}. Some recent neutron scattering results \cite{Bourges} have been interpreted in terms of the Varma phase with circulating currents \cite{Varma}. Nevertheless, no consensus has been achieved yet among the different scenarios proposed theoretically for explaining the pseudogap phase \cite{Varma,Rome,Vojta}
and an unambiguous answer to this puzzle is as yet lacking. 

An experimental breakthrough in the YBCO family of cuprates was the measurement of the electronic structure via quantum oscillations, thereby enabling access to low energy coherent quasiparticles. The measurement of Shubnikov de Haas and de Haas van Alphen oscillations in YBa$_2$Cu$_3$O$_{6+\delta}$ \cite{Leyraud1,Jaudet1,Sebastian2} and YBa$_2$Cu$_4$O$_8$ \cite{Yelland1,Bangura1} was made possible by advances involving high magnetic fields and improved single crystal quality. At high magnetic fields where superconductivity is destroyed, a Fermi surface comprising small sections was measured in these underdoped cuprate materials, indicating likely translational symmetry breaking that `reconstructs' the large paramagnetic Fermi surface. It is currently a subject of debate and ongoing experiments as to the origin of such a superlattice and its relation to unconventional superconductivity. Another outstanding question relates to the apparent dichotomy between `Fermi arcs' measured at high temperatures and low magnetic fields by photoemission experiments \cite{Kanigel1,Tanaka1}, and the closed `Fermi pockets' measured at low temperatures and high magnetic fields by quantum oscillation measurements. Experiments that relate these two regimes are crucial to understand quasiparticle excitations in the precursor phase to superconductivity, thereby providing potential clues as to the Cooper pairing mechanism.

\section{UNCONVENTIONAL SUPERCONDUCTIVITY: SAME OR DIFFERENT}
It is striking that the occurrence of superconductivity on the brink of magnetisation appears to be ubiquitous. Indeed, common elements are in play in the assorted $f$- and $d$- electron families discussed: a low-dimensional crystal structure, (anti)ferromagnetism and tuneability to the brink of magnetism. Similar tuning parameters such as doping and applied pressure are seen to suppress magnetism and induce superconductivity in these materials, the particular attraction of the $d$-electron family of materials lying in their significantly higher energy scales. 

The commonality in behaviour which becomes apparent on considering representative materials' families, however, poses a puzzle. On further inspection, this apparently universal behaviour is found to display surprising variations in details of mediating mechanisms and mediated phases. This deviation from notions of universality leads us to another important consideration that may inform such dichotomous behaviour - the flattened energy landscape in the vicinity of a phase instability. The consequent degeneracy of phases in this region needs to be weighed in the balance with potential universal behaviour in order to understand the similar yet different manifestation of physical phenomena.

\section{ULTRA STRONGLY CORRELATED PHENOMENA IN GRAPHITE AND BISMUTH}

While unconventional superconductivity is a consequence of strong correlations in condensed matter systems, in the limit of ultra strong correlations, arguably more exotic effects come into play. A 2D electron gas in the presence of a perpendicular magnetic field offers a rich playground for the observation of such exotic quantum states of matter, celebrated examples of which include the fractional QHE states such as the Abelian Lauglin liquid \cite{Laughlin} or the non-Abelian Pfaffian and parafermionic states \cite{MR,RR}. At this year's SCES, it was suggested that practical 3D materials, examples being graphite and bismuth, may also exhibit such quantum effects.

\subsection{Graphite}

Although graphite is a 3D material consisting of several graphene planes, the high-anisotropy between out- and in-plane transport observed in highly oriented pyrolytic graphite (HOPG) $\rho_{out} / \rho_{in} \sim 5.10^4$ \cite{1} puts this material in the class of 2D conductors. In the presence of a perpendicular magnetic field, 2D electron systems are expected to display QHE. Integer QHE in the presence of a perpendicular magnetic field has been previously demonstrated in HOPG \cite{1}. Careful analysis of the steps in HOPG further showed that this material exhibits QHE for both Dirac-like holes and massive electrons \cite{2}. This observation that the Fermi surface in graphite comprises both electron and hole pockets is also consistent with de Haas van Alphen and Shubnikov de Haas quantum oscillations \cite{3}, scanning tunneling spectroscopy \cite{4}, far-infrared magneto-transmission spectroscopy \cite{5}, and angle-resolved photoemission experiments \cite{6}.

At this year's SCES, studies were presented on HOPG with improved mobility ($\mu \sim 10^6 $ cm$^2$/Vs) and ultra-high pulsed magnetic fields (up to $B = 57$T) applied perpendicular to the graphene layers \cite{7}. Deep in the quantum limit, for fields $B \gg B_{QL} \sim 7-8$T, several plateaus are seen in the Hall resistivity at fractional filling factors. Although the longitudinal resistivity $\rho_{xx}$ does not vanish in the plateau region, it exhibits small dips at the filling factors for which $\rho_{xy}$ shows plateaus, as do other 3D systems. The series of plateaus observed at fractional filling factors $\nu = 2/7, 1/4, 1/5, 2/11, 1/6, 1/8, 2/17$, and $1/9$ \cite{7} suggests fractional QHE in graphite, albeit a more complex scenario than the Jain series that appears in more conventional 2D electron system \cite{8}.

\subsection{Bismuth} 

A rather more surprising example of a 3D material where fractional QHE is potentially realised is bismuth \cite{9}. Due to the extremely small Fermi surface of Bismuth and the long Fermi wavelength of itinerant electrons in this system, the quantum limit can be attained by applying moderate magnetic fields $B \sim 9$T along the trigonal axis of the material \cite{10}.  Earlier studies up to $B = 12$T showed quantum oscillations in the Nernst coefficient in the vicinity of the quantum limit \cite{11}. It was further argued that the peak in the Nernst signal where the Landau level crosses the Fermi level is indicative of a `quantum Nernst effect' associated with integer QHE.

At this year's SCES, measurements of transport properties of a single crystal of bismuth under magnetic fields up to $B = 35$T were reported \cite{9}, revealing a plateau in the Hall resistivity at a {\it fractional} filling factor. The previous detection of three peaks in the Nernst response corresponding to rational fractions 2/3, 2/5, and 2/7 of the first integer peak were the first indications of a fractional QHE in this material \cite{10}. However, no plateaus were observed at the corresponding Hall resistivity data, which had only low resolution at $T = 0.44$K  \cite{10}. Recently, a clear plateau-like feature was measured for magnetic fields applied at magic angles with respect to the trigonal axis \cite{9}, at a filling factor which would naively correspond to $\nu = 1/3$ for holes. While the high mobility and small Fermi surface of bismuth make it a promising candidate for exotic quantum effects such as fractional QHE, it remains a puzzle as to how its 3D structure could support such a state.

These novel observations may introduce more questions than answers. For instance, the involvement of the rich Fermi surfaces in graphite and bismuth systems that comprise both electron and hole pockets is as yet unclear. Whereas in graphite the electrons are massive and the holes are Dirac-like, in bismuth the situation is the opposite - the holes are massive and the electrons are massless - it is conceivable that competing responses from electrons and holes introduce novel effects. It is hoped that further investigations of higher mobility samples at lower temperatures and stronger magnetic fields could shed further light in this direction.

\section{COLD ATOMS}

The experimental realization of ultracold atomic gases loaded into optical lattices has opened a unique pathway to study strongly correlated quantum many-body systems. The great versatility in engineering  optical lattices allows for a full control of the lattice parameter and of
the barrier (height and width) separating neighboring sites. In addition, the lattice can be loaded with bosons, fermions, or Bose-Fermi mixtures and the inter- and/or intra-species interactions can be tuned from attractive to repulsive by using the technique of Feshbach resonance \cite{Feshbach}. 
Several interesting phenomena predicted in the context of condensed matter systems have recently been observed with cold atoms. A prominent example is the detection of the superfluid/Mott insulator transition in 3D \cite{Bloch} and more recently in 2D \cite{Phillips}
optical lattices loaded with bosons. In the limit of strong repulsive on-site interactions $U$ the bosons are localized and form a Mott insulator, whereas a superfluid phase with spontaneously broken gauge symmetry emerges below the critical value of the ratio between $U$ and the 
hopping parameter $t$. Other interesting examples are the observation of the crossover between the Bose-Einstein condensate and the BCS regimes by varying the interaction strength in fermionic condensates \cite{BEC-BCS}, and more recently the creation of an antiferromagnetic Neel state
with spinful fermionic atoms \cite{BlochAF}. 

In this context, some novel realizations of strongly correlated phases were proposed during this year's SCES. One of them addressed the investigation of Luttinger liquid physics by generating 1D tubes of strongly interacting fermionic isotopes in a 2D optical lattice. If the tubes are
well separated, two regimes are expected to occur, depending on the total atomic density and on the 3D s-wave scattering length between different species of atoms. In the Spin-Coherent regime, the usual spin-charge separation is expected to occur. However, another regime can
be realized, in which the Luttinger liquid becomes `Spin-Incoherent' and only charge excitations remain as a collective mode. The measurement of the off-diagonal (spin-up spin-down) correlator of density fluctuations was proposed to be the natural observable for detecting both phases, since it would exhibit different signs in the two different regimes \cite{Bolech}. 

Another proposal considered the effect of a staggered rotation for atoms loaded in a 2D square optical lattice \cite{Andi}. The rotation acts as an effective staggered magnetic field and the Hamiltonian describing the system becomes a generalized Hubbard model, with complex and anisotropic hopping 
coefficients \cite{Lih08}. For the case of bosons, the system exhibits different superfluid phases at small values of the Hubbard parameter $U$, depending on the strength of the `gauge' field. For weak magnetic fields, the bosons condense at zero momentum and the conventional uniform superfluid is realized.
This phase is reminiscent of the Meissner phase in superconductors. On the other hand, for stronger fields the minimum of the single-particle spectrum moves to the boundary of the Brillouin zone and a finite-momentum condensate is realized \cite{Lih08}. This phase bears analogies with the FFLO state; nevertheless, 
in the FFLO the Cooper pairs carry a finite linear momentum, whereas in this case the bosons carry a finite angular momentum. Indeed, this phase consists of a square vortex-antivortex lattice, analogous to the Abrikosov-vortex lattice observed in type-II superconductors. The quantum phase transition between
the uniform and the finite momentum condensates occurs when the magnetic flux per plaquette equals one half of the fundamental flux-quantum $\phi_0$ and it is of first order. 

When loaded with
fermions, instead, at half filling this system realizes the physics of graphene \cite{Antonio}. Due to the staggered rotation, the optical lattice is divided into A and B sublattices and the single-particle spectrum exhibits 4 Dirac cones, with two inequivalent ones.  At the critical value of the `magnetic field' where 
the flux $\phi = \phi_0 / 2$ the fermionic system exactly realizes the graphene spectrum because the Dirac cones are isotropic. However, by changing the magnetic field the cones become anisotropic in the $x$ and $y$ directions and the cold atom system shares the features proposed to occur by depositing graphene on top of a periodic potential \cite{Park}. This superlattice configuration could be very important for technological applications involving graphene. Moreover, at this critical field the fermionic
system realizes the so called staggered-$\pi$ flux phase, which was proposed long ago to be the
ground state in the pseudogap phase of high-$T_c$ cuprates \cite{Marston}.

An even more interesting case emerges when this staggered lattice is loaded with both, fermions and
bosons. In this case an unconventional superconducting phase  can arise because a nearest-neighbor attractive interaction between fermions in the A and B sublattices can be generated, mediated
by the bosons. Due to the additional sublattice degree of freedom, a state which is singlet in the spin and in the sublattice but odd in the orbital can occur, thus opening the possibility to realize unconventional superconductivity with cold atoms \cite{Lih09}. From the theoretical point of view, it is by now clear that cold atom systems under staggered rotation in an optical lattice offer the possibility of
realizing a panoply of interesting quantum states of matter, ranging from finite-momentum Bose-Einstein condensates, to anisotropic Dirac fermions, or even unconventional superconductivity. The experimental realization of such exotic phases remains a challenge to be faced in the forthcoming years.  

\section{WHERE TO LOOK NEXT}

A veritable bevy of experiments is currently underway to unearth details of the nature of superconductivity, magnetism, and possible interaction mechanisms that might potentially mediate unconventional superconductivity in systems such as heavy fermions, cuprate and pnictide high $T_c$'s. Amidst all this busyness, however, a pressing question to be asked is: why was potential high temperature superconductivity in the pnictides not explored earlier, given the shared characteristics that appear evident on hindsight? As we have seen, diverse and unexpected phenomena in disparate materials families can nonetheless be woven into the narrative of novel phases nucleated at a near-continuous phase instability. Candidates with at least nominally similar characteristics: crystal structure and magnetisation for instance, would be the natural candidates to search for unconventional phenomena, and not just superconductivity - but possibly even more exotic pairing symmetries, for instance spontaneous forms of current order.

A further boost in this effort to identify candidates for novel quantum phases may be provided through cold atom models. These models hold the promise of tailoring, under extremely well controlled conditions, the most exotic correlated systems. Such an initiative could play a pivotal role in the understanding of the interplay between magnetism and superconductivity in these complex systems. 

While various discoveries of novel phases can be collectively viewed through the prism of quantum critical phenomena, adherence to this prescription is likely to yield not a blandly universal array of physical phenomena, rather a rich landscape of novel phases and mediating mechanisms peculiar to each materials system studied. Looking forward toward developing a condensed matter community that seeks to explore new and uncharted territory, it cannot be sufficiently emphasised as to the importance of a directed search for and study of candidate new materials families.


\begin{thebibliography}{}

\bibitem{Hertz1} J. A. Hertz, Phys. Rev. B 14 (1976) 1165.
\bibitem{Millis1} A. J. Millis, Phys. Rev. B 48 (1993) 7183.
\bibitem{Bucher1} E. Bucher et al., Phys. Rev. B. 11 (1974) 440.
\bibitem{Ott1} H. R. Ott et al., Phys. Rev. Lett. 50 (1983) 1595.
\bibitem{Steglich1} F. Steglich et al., Phys. Rev. Lett. 43 (1979) 1892.
\bibitem{Petrovic1} C. Petrovic et al., J. Phys.: Condens. Matter 13 (2001) L337.
\bibitem{Mathur1} N. D. Mathur et al., Nature 394 (1998) 39.   
\bibitem{Larkin1} A. L. Larkin, Y. N. Ovchinnikov, Sov. Phys. JETP 20 (1965) 762.
\bibitem{Fulde1} P. Fulde, R. A. Ferrell, Phys. Rev. A. 550 (1964) 135.
\bibitem{Matsuda1} Y. Matsuda, H. Shimahara, J. Phys. Soc. Jpn. 76 (2007) 051005, and references therein.
\bibitem{Bianch1} A. D. Bianchi et al., Science 319 (2008) 177.
\bibitem{Nicklas1} M. Nicklas et al., J. Low Temp. Phys., 146 (2007)
\bibitem {Kenzelmann1} M. Kenzelmann et al., Science 321 (2008) 1652.
\bibitem{Sarrao1} J.L. Sarrao et al., Nature, 420, (2002) 297.
\bibitem{Curro1} N. J. Curro et al., Nature 434 (2005), 622.
\bibitem{Caicuffo1} F. Jutier et al., Phys. Rev. B 77 (2008) 024521.
\bibitem{Flint1} R. Flint, M. Dzero, P. Coleman, Nat. Phys. 4 (2008), 643.
\bibitem{Nakatsuji1} S. Nakatsuji et al., Nat. Phys. 4 (2008), 603.
\bibitem {OFarrell} E. C. T. O'Farrell et al., arXiv:0811.4417v1 (2008).
\bibitem{Aoki1} D. Aoki et al., J. Phys. Soc. Jpn. 76 (2007) 063701.
\bibitem{Saxena1} S. S. Saxena et al., Nature 406 (2000) 587.
\bibitem{Aoki2} D. Aoki et al., Nature 413 (2001) 613.
\bibitem{Levy1} F. L$\acute{e}$vy et al., Science 309 (2005) 1343.
\bibitem{Levy2} F. L$\acute{e}$vy, th$\grave{e}$se (2006).
\bibitem{Miyake1} A. Miyake, D. Aoki, J. Flouquet, J. Phys. Soc. Jpn. 77 (2008) 004709.
\bibitem{Levy3} F. L$\acute{e}$vy, I. Sheikin, A. Huxley, Nat. Phys. 3 (2007) 460.
\bibitem{Huy1} N. T. Huy et al., Phys. Rev. Lett. 99 (2007) 067006.
\bibitem{Bruning1} E. M. Br$\ddot{a}$ning et al., Phys. Rev. Lett., 101 (2008) 117206.
\bibitem{Kamihara1} Y. Kamihara et al., J. Am. Chem. Soc. 128 (2006) 10012.
\bibitem{Tegel1} M. Tegel, D. Bichler, D. Johrendt, Solid State Sciences 10 (2008) 193.
\bibitem{Kamihara2} Y. Kamihara et al., J. Am. Chem. Soc. 130 (2008) 3296.
\bibitem{Norman1} M. R. Norman, Physics 1 (2008) 21 and references therein.
\bibitem{Ren1} Z.-A. Ren, et al., Chin. Phys. Lett. 25 (2008) 2215.
\bibitem{Hu1} W. Z. Hu et al., Phys. Rev. Lett. 101 (2008) 257005.
\bibitem{Sebastian1} S. E. Sebastian et al., J. Phys.: Condens. Matter 20 (2008) 422203.
\bibitem{Uemura1} Y. J. Uemura, arXiv:0811.1546v1 (2008).
\bibitem{Yazdani}  K. K. Gomes et al, Nature 447 (2007) 569. 
\bibitem{Bourges} Y. Li et al, Nature 455 (2008) 372. 
\bibitem{Varma} C. Varma, Phys. Rev. B 55 (1997) 14554.; 73 (2006) 155113. 
\bibitem{Rome} M. Grilli et al., arXiv:0903.2588; 
\bibitem{Vojta} M. Vojta, arXiv:0901.3145. 
\bibitem{Leyraud1} N. Doiron-Leyraud et al., Nature 447 (2007) 565. 
\bibitem{Jaudet1} C. Jaudet et al., Phys. Rev. Lett. 100 (2008) 187005.
\bibitem{Sebastian2} S. E. Sebastian et al., Nature 454 (2008) 200.
\bibitem{Yelland1} E. A. Yelland et al., Phys. Rev. Lett. 100 (2008) 047003.
\bibitem{Bangura1} A. F. Bangura et al., Phys. Rev. Lett. 100 (2008) 047004.
\bibitem{Kanigel1} A. Kanigel et al., Nat. Phys. 2 (2006) 447.
\bibitem{Tanaka1} K. Tanaka et al., Science 314 (2006) 1910.
\bibitem{Laughlin} R. B. Laughlin, Phys. Rev. B 23 (1981) 5652; Phys. Rev. Lett. 51 (1983) 605. 
\bibitem{MR} G. Moore and N. Read, Nucl. Phys. B 360 (1991) 362. 
\bibitem{RR} N. Read and E. H. Rezayi, Phsy. Rev. B 59 (1999) 8084. 
\bibitem{1} Y. Kopelevich et al., Phys. Rev. Lett. 90 (2003) 156402.
\bibitem{2}I. A. Luk'yanchuk and Y. Kopelevich, Phys. Rev. Lett. 97 (2006) 256801.
\bibitem{3} I. A. Luk'yanchuk and Y. Kopelevich, Phys. Rev. Lett. 93 (2004) 166402.
\bibitem{4} G. Li and E.Y. Andrei, Nature Phys. 3 (2007) 623.
\bibitem{5} M. Orlita, C. Faugeras, and G. Martinez, Phys. Rev. Lett. 100 (2008) 136403.
\bibitem{6} S.Y. Zhou, G.-H. Gweon, and A. Lanzara, Annals of Phys. 321 (2006) 1730; Nature Phys. 2 (2006) 595.
\bibitem{7} Y. Kopelevich et al., unpublished (2009). 
\bibitem{8} J.K. Jain, Phys. Rev. Lett. 63 (1989) 199.
\bibitem{9} B. Fauqu\'e, H. Yang, I. Sheikin, L. Balicas, J.-P. Issi, and K. Behnia, arXiv:0902.3103.
\bibitem{10} K. Behnia, L. Balicas, and Y. Kopelevich, Science 317 (2007) 1729.
\bibitem{11} K. Behnia, M.-A. M\'easson, and Y. Kopelevich, Phys. Rev. Lett. 98 (2007) 166602.
\bibitem{12} C. Biagini, D.L. Maslov, M.Yu Reizer, and L.I.Glazman, Europhys. Lett. 55 (2001) 383.
\bibitem{Feshbach} H. Feshbach, Ann. Phys. 5 (1958) 357. 
\bibitem{Bloch} M. Greiner et al., Nature (London) 415 (2002) 39. 
\bibitem{Phillips} I. B. Spielman, W. D. Phillips, and J. V. Porto, Phys. Rev. Lett. 98 (2007) 080404. 
\bibitem{BEC-BCS} I. Bloch, J. Dalibard, and W. Zwerger, Rev. Mod. Phys. 80 (2008) 885. 
\bibitem{BlochAF} U. Schneider et al., Science 322 (2008) 1520.

\bibitem{Bolech} P. Kakashvili, S. G. Bhongale, H. Pu, and C. Bolech, Phys. Rev. A 78 (2008) 041602(R). 
\bibitem{Andi} A. Hemmerich and C. Morais Smith, Phys. Rev. Lett. 99 (2007) 113002.
\bibitem{Lih08} Lih-King Lim, C. Morais Smith, and Andreas Hemmerich, Phys. Rev. Lett. 100 (2008) 130402.
\bibitem{Antonio} A. H. Castro Neto et al., Rev. Mod. Phys. 81 (2009) 109. 
\bibitem{Park} C.-H. Park, L. Yang, Y.-W. Son, M. L. Cohen, and S. G. Louie, Nature Physics 4 (2008) 213. 
\bibitem{Marston} I. Affleck and J. B. Marston, Phys. Rev. B 37 (1988) 3774; J. B. Marston and 
I. Affleck, Phys. Rev. B 39 (1989) 11 538. 
\bibitem{Lih09} Lih-King Lim, A. Lazarides, A. Hemmerich, and C. Morais Smith, unpublished (2009). 

\end{thebibliography}
\end{document}